# Conversion of recoilless γ-radiation into a periodic sequence of ultrashort pulses in a set of dispersive and absorptive resonant media

Y.V. Radeonychev[1,2,3,*], V.A. Antonov[1,2,3], F.G. Vagizov[3,5], R. N. Shakhmuratov[3,4] and Olga Kocharovskaya[5]

[1] *Institute of Applied Physics, Russian Academy of Sciences,
46 Ulyanov Street, Nizhny Novgorod, 603950, Russia,*
[2] *N.I. Lobachevsky State University of Nizhny Novgorod,
23 Gagarin Avenue, Nizhny Novgorod, 603950, Russia,*
[3] *Kazan Federal University, 18 Kremlyovskaya Street,
Kazan 420008, Republic of Tatarstan, Russia,*
[4] *Kazan Physical-Technical Institute, Russian Academy of Sciences,
10/7 Sibirsky Trakt, Kazan 420029 Russia*
[5] *Department of Physics and Astronomy and
Institute for Quantum Studies and Engineering,
Texas A&M University, College Station, TX 77843-4242, USA.*

**Abstract** An efficient technique to produce a periodic sequence of ultrashort pulses of recoilless γ-radiation via its transmission through the optically thick vibrating resonant absorber was demonstrated recently [Nature, 508, 80 (2014)]. In this work we extend the theoretical analysis to the case of a set of multiple absorbers. We consider an analytical model describing the control of spectral content of a frequency modulated γ-radiation by selective correction of amplitudes and initial phases of some spectral components, using, respectively, the resonant absorption or dispersion of nuclei. On the basis of the analytical solutions we determine the ultimate possibilities of the proposed technique.

## I. INTRODUCTION

In recent years, there has been an active search for methods to generate and control electromagnetic radiation with photon energy from units to tens of kiloelectronvolts (keV) [1-16]. This search is motivated by the potential to significantly extend the existing diagnostic capabilities, including the high-precision spatial measurements [4] and structural time-resolved studies of fast physical, chemical and biological processes [6-14]. Shortening the wavelength reduces the diffraction-limited focus spot, which opens up the novel possibilities in high-precision lithography [15] and focusing the radiation to extremely large intensity [16]. Novel opportunities in quantum optics emerge in X-ray domain [4]. One of the main directions of research in this area is to adapt the concepts of nonlinear and coherent optics to the keV photon energy range [1-13].

In the frequency domain covering the photon energies of tens of keV the nonresonant methods of nonlinear and coherent optics to control radiation are rather limited because of strong ionization of atoms. At the same time, photons of such energies can effectively interact with nuclei. Of particular importance here are the processes of recoilless resonant interaction of photons with nuclei based on Mössbauer effect [17-37]. Several analogues of the well-known optical effects were observed in the resonant recoilless interaction of gamma-photons with nuclei, such as the ac-Stark splitting [23], so-called gamma-echo effect [24-26], magnetic switching of nuclear forward scattering [28-30], the collective Lamb shift [31], cavity electromagnetically induced transparency [32], slowing down γ-photons in a doublet structure of the nuclear transition [21], single-photon revival in nuclear absorbing multilayer structures [33], and vacuum-assisted generation of coherences [34].

Most recently, the efficient converter of Mössbauer radiation of a radioactive source into a periodic sequence of ultrashort pulses was demonstrated [35]. The converter constitutes a stainless steel foil with $^{57}$Fe nuclei placed in the path of propagation of 14.4keV recoilless radiation emitted by a radioactive source $^{57}$Co. The foil harmonically vibrates along the propagation direction of radiation. The spectral and temporal conversion of radiation occurs as a result of its resonant interaction with $^{57}$Fe nuclei during the passage through the optically thick foil. Tuning the frequency and amplitude of vibration, resonance energies of the emitting and absorbing nuclei, and thickness of the foil enables one to control the shape, duration and repetition period of the produced pulses. The pulses with minimum duration of 18 ns were obtained in [35]. Such a technique is based on the quite general approach to control the spectral-temporal properties of radiation in various frequency ranges [38-48]. In the γ-ray domain the pulse shaping is successful owing to the following three basic factors. (i) Very short wavelength of radiation (≈0.96Å for 14.4keV transition of $^{57}$Co) and uniquely large ratio of resonance energy of the nuclei to linewidth (≈3×10$^{12}$ for 14.4keV transition of both $^{57}$Co and $^{57}$Fe) makes effective using the Doppler effect in the photon-nuclei interaction to control the resonance energies of both the source and converter. (ii) Harmonic Doppler modulation of the resonance energy of the converter with respect to the incident photon energy via vibration enables one to generate a wide spectrum of radiation sidebands with predetermined initial phases. (iii) Resonant alteration of the selected spectral components of the photons inside the converter makes possible to build the phase aligned spectrum corresponding to the pulsed waveforms of radiation.

In this paper, we further develop the above technique, extending it to the case of the converter constituting a set of multiple resonant absorbers. We consider an analytical model describing the control of spectral content of a frequency modulated γ-radiation by selective correction of amplitudes and initial phases of some spectral components, using, respectively, the resonant absorption or dispersion of nuclei. We discuss the possibilities to shorten the pulses and to increase their maximum peak intensity as well as to form more complicated temporal waveforms utilizing the resonant recoilless media.

The paper is organized as follows. In section II we describe the theoretical model, in section III we study analytically the techniques to build the phase matched spectra corresponding to various temporal waveforms based on the deletion of a single or several components from the spectrum of the frequency-modulated radiation via their attenuation in a single or several sequentially placed resonant absorbers. In section IV we analytically investigate the techniques to build the phase aligned spectra corresponding to periodic sequences of pulses with higher peak intensity and shorter duration based on inversion of initial phases of the mismatched components of the spectrum of frequency-modulated radiation in properly tuned resonant absorbers. In section V we summarize the results.

## II. THE MODEL

Our study is based on the analytical solutions of the following model. The photons emitted by a radioactive source form a flow of γ-radiation. Taking into account the extreme narrowness of recoilless spectral line of the source, we consider the emitted radiation as a monochromatic wave. In this case the electric field of the emitted radiation, $E_{em}$, can be written as

$$E_{em}(\tau) = E_0 e^{-i\omega_0 \tau}. \tag{1}$$

Here $E_0$ is the amplitude of electric field determined as $E_0 = \sqrt{8\pi I_0/c}$, where $I_0$ is the intensity of radiation experimentally found as $I_0 = N\hbar\omega_0/s$ (where $\hbar\omega_0$ is the energy of photon emitted by the source, and $N$ is the average number of photons registered per unit time per unit square $s$), and $\tau = t - z/c$ is the local time in the reference frame co-moving with γ-ray photons along "$z$"-direction with the speed of light in vacuum $c$. The field (1) propagates through a resonant absorber, which is a layer of a medium containing recoilless nuclei with quantum transition near-resonant to the field.

The photon energy, $\hbar\omega_0$, can be made harmonically modulated with respect to the energy of quantum transitions of the absorber nuclei by two ways. The first way is to fix the absorber and vibrate the source as a whole along the propagation direction of the field (1), such that coordinate $z$ in the laboratory reference frame is related to the corresponding coordinate $z_s$ in the reference frame associated with nuclei of the vibrating source, as

$$z = z_s - R\sin(\Omega t + \vartheta_0),  \qquad (2)$$

where $R$, $\Omega$, and $\vartheta_0$ are the amplitude, angular frequency, and initial phase of vibration, respectively. This relation is valid if thickness of the source, $L_s$, meets inequality $L_s \ll 2\pi V_{sound}^s/\Omega$, where $V_{sound}^s$ is the speed of sound inside the source. The field (1) emitted by the vibrating source takes, because of the Doppler effect, the form of the frequency-modulated wave in the laboratory reference frame [19],

$$E_{em}(\vartheta_0,\tau) = E_0 e^{-i(\omega_0\tau - P\sin(\Omega\tau+\vartheta_0))} \qquad (3)$$

where $P = R\omega_0/c$ is the modulation index.

The second way is to fix the source and vibrate the absorber as a whole along the propagation direction of the field (1), such that coordinate $z$ in the laboratory reference frame is related to the corresponding coordinate $z_a$ in the reference frame associated with nuclei of the moving absorber, as

$$z = z_a + R\sin(\Omega t + \vartheta_0). \qquad (4)$$

Similarly, thickness of the absorber, $L_A$, should meet the relation $L_A \ll 2\pi V_{sound}^A/\Omega$, where $V_{sound}^A$ is the speed of sound inside the absorber. Being monochromatic in the laboratory reference frame, field (1) takes the form (3) (where $\tau = t - z_a/c$) in the reference frame of the vibrating absorber.

The spectrum of the frequency-modulated field (3) consists of discrete components separated by the vibration frequency,

$$E_{em}(\vartheta_0,\tau) = E_0 e^{-i\omega_0\tau} \sum_{m=-\infty}^{\infty} J_m(P) e^{im(\Omega\tau+\vartheta_0)} = \sum_{n=-\infty}^{\infty} \tilde{E}_n e^{-i\omega_n\tau} = \sum_{n=-\infty}^{\infty} E_n(\vartheta_0,\tau) \qquad (5)$$

where $\omega_n = \omega_0 + n\Omega$ and $\tilde{E}_n = E_0 J_{-n}(P) e^{-in\vartheta_0}$ are the frequency and complex amplitude of the $n$-th spectral component, and $J_{-n}(P)$ is the Bessel function of the first kind meeting the relation $J_{-n} = e^{-i\pi n} J_n$. The amplitudes of spectral components are determined by the amplitude of vibration via the Bessel functions, $|J_n(P)|$. The initial phases of spectral components are either $0 - n\vartheta_0$ (let us call them in-phase components) if $J_{-n}(P) > 0$ or $\pi - n\vartheta_0$ (let us call them antiphase components) if $J_{-n}(P) < 0$. The representation (3), (5) is valid in the case where the modulation frequency is much larger than the spectral linewidth of the source [49].

Let us consider the possible techniques to convert the frequency-modulated field (3), (5) with time-independent amplitude, $E_0$, into an amplitude-modulated field having a form of a periodic sequence of ultrashort γ-ray pulses with the peak height noticeably exceeding the amplitude $E_0$. It is known that a periodic sequence of pulses of minimum duration corresponds to the widest spectrum of equidistant perfectly synchronized frequency components. In the best case the phase difference between the neighboring spectral components is constant value over the whole spectrum, i.e. the components are phase aligned. The corresponding pulses, called bandwidth limited pulses, have minimum possible duration while the bursts between the pulses have the smallest height. If, in addition, all the frequency components have the same amplitude, the peak intensity of the pulses achieves the maximum limit whereas the pedestal vanishes. The equidistant spectrum (5) of the frequency-modulated field (3) can be very wide in the case of large value of the modulation index, $P$ (the spectrum width is normally evaluated as $\Delta\omega = 2(P+1)\Omega$). However, the amplitudes and initial phases of the spectral components have the specific magnitudes such that superposition of the corresponding monochromatic waves results in the field with constant amplitude, $|E(\tau)| = E_0$. Therefore, in order to convert the field (3) into a pulse train, one needs to correct amplitudes and phases of the spectral components. As noted above, the nonresonant methods of nonlinear optics to align phases of the spectral components fail to be implemented in a keV-domain because of strong attenuation of radiation due to photoionization [19]. Let us consider how the resonant properties of recoilless absorbing nuclei can be utilized to build a phase aligned spectrum of Mössbauer radiation corresponding to a periodic sequence of pulses.

Let the field (3) propagates through one or several absorbers having spectral lines of the resonant transitions much narrower than the frequency interval between the spectral components. Tuning the spectral lines of the absorbers to the frequencies of the chosen spectral components results in selective alteration of these components. Selective alteration of the complex amplitudes of $N$ spectral components with numbers $m = \underbrace{a,b,c,...}_{N}$ of the spectrum (5) due to the resonant absorption and phase incursion can be described by the relation

$$E_{tr}(a,b,...,\vartheta_0,\tau) = E_{em}(\vartheta_0,\tau) \underbrace{-(1-\tilde{\eta}_a)\tilde{E}_a e^{-i\omega_a \tau} - (1-\tilde{\eta}_b)\tilde{E}_b e^{-i\omega_b \tau} - ...}_{N}, \quad (6)$$

where $E_{tr}(a,b,...,\vartheta_0,\tau)$ is the total transformed field after alteration of spectral components with numbers $a, b, ...$ and the coefficient $\tilde{\eta}_m = e^{-\Lambda_m/2}e^{-i\Phi_m}$ (where $m=a,b, ...$) describes the decrease of amplitude and incursion of phase of $m$-th spectral component. The selective modification of spectral components inside the resonant absorbers results in amplitude modulation of the transmitted field intensity, $I_{tr} = c|E_{tr}|^2/8\pi$. Below we discuss the cases, where such a modulation constitutes periodic sequences of pulses.

## III. DELETION OF SPECTRAL COMPONENTS

Let us consider deletion of a single or several spectral components via their selective attenuation in the properly tuned single or several resonant absorbers. Let $N$ components with numbers $m = \underbrace{a,b,c,...}_{N}$ are deleted from the spectrum (5). Then one has $\tilde{\eta}_m = 0$ in (6) and the resulting field is

$$E_{del}(a,b,...,\vartheta_0,\tau) = E_{em}(\vartheta_0,\tau) \underbrace{-\tilde{E}_a e^{-i\omega_a \tau} - \tilde{E}_b e^{-i\omega_b \tau} - ...}_{N}. \quad (7)$$

Its intensity, $I_{del}=c|E_{del}|^2/8\pi$, registered by the detector behind the absorbers, can be written in the form:

$$I_{del}(a,b,...,\vartheta_0,\tau) = I_0 \Big\{ \underbrace{1 + J_{-a}^2 + J_{-b}^2 + ...}_{N}$$
$$\underbrace{-2J_{-a}\cos(P\sin\alpha + a\alpha) - 2J_{-b}\cos(P\sin\alpha + b\alpha) - ...}_{N} \quad (8)$$
$$\underbrace{+ 2J_{-a}J_{-b}\cos((a-b)\alpha) + 2J_{-a}J_{-c}\cos((a-c)\alpha) + 2J_{-c}J_{-b}\cos((c-b)\alpha) + ...}_{C_N^2} \Big\}$$

where $I_0 = cE_0^2/8\pi$ is the intensity of radiation emitted by the source prior to any transformation, and $\alpha = \Omega\tau + \vartheta_0$.

Note that, according to (8), removing an arbitrary set of $N$ components with numbers $\underbrace{a,b,c,...}_{N}$ from the spectrum (5) of the frequency-modulated field (3) results in the same temporal dependence of the intensity, as removing the set of N components with numbers $\underbrace{-a, -b, -c, ...}_{N}$ (symmetrically located with respect to the carrier frequency $\omega_0$) subject to the initial phase $\vartheta_0$ is shifted by $\pi$. In other words, the following relation is valid,

$$I_{del}(a, b, ..., \vartheta_0, \tau) = I_{del}(-a, -b, ..., \vartheta_0 \pm \pi, \tau). \tag{9}$$

The initial phase, $\vartheta_0$, sets the time, $\tau_0$, when the phase of vibration of the source or absorber equals zero. Its change results in shift of the temporal pattern of intensity as a whole in time. Below we assume that $\vartheta_0 = 0$.

As follows from (8), the intensity of the field passed through the absorbers, $I_{del}$, can periodically with period $2\pi/\Omega$ exceed the intensity of the field emitted by the source. The fist peak of the intensity is achieved at $\tau=0$,

$$I_{del}(a, b, ..., \tau = 0) = I_0 \left(1 \underbrace{- J_{-a} - J_{-b} ...}_{N}\right)^2, \tag{10}$$

provided that $J_{-n}<0$ for $n=a,b,...$. This corresponds to the deletion of spectral components with initial phases of $\pi$ (antiphase spectral components). The deletion of all antiphase components from the spectrum (5) would lead to formation of a train of bandwidth limited pulses. The peak pulse intensity takes maximum value, $I_{del}(a, b, ..., \tau = 0) = I_{del}^{max}(a, b, ...)$, if the modulation index, $P$, provides maximum magnitude of the sum $\underbrace{|J_{-a}| + |J_{-b}| + ...}_{N}$. This can be explained as follows. In order to provide the constant amplitude of the field (3), the antiphase spectral components compensate the in-phase components. Deletion of the antiphase components results in destruction of this balance. The larger total amplitude of the deleted components, the stronger disbalance takes place.

Let us study the conditions when the spectrally selective filtering of the field (3) in a single or several resonant absorbers leads to pulse sequence with minimum pulse duration and maximum pulse height. If the modulation index $P<2.4$, the spectrum (5) consists of the central component at carrier frequency, $\omega_0$, and no more than four significant sidebands of positive and four sidebands of negative orders, $n$, according to the magnitudes of the correspondnig Bessel functions (Fig. 1). Two spectral components with indices $a=1$, $b=3$ are antiphase. In the case of $P=1.84$, realized in the proof-of-principle experiment [35], the amplitude of the third spectral component is small enough ($-J_{-3}(1.84)=0.1$, see Fig.1, and Fig. 2 inset, black spectrum #1) and can be neglected. Then, according to (10), the maximum peak pulse intensity is achieved when the first component is deleted subject to $P=1.84$, namely $-J_{-1}(1.84) = 0.58 = $ Max corresponding to $I_{del}^{max}(1) = 2.5 I_0$ (Fig. 2, black line #1). At larger modulation index, $P=2.3$, amplitude of the third spectral sideband is $-J_{-3}(2.3)=0.18$ while the sum $-J_{-3} - J_{-1}$ takes its maximum value, i.e., $-J_{-3}(2.3) - J_{-1}(2.3) = 0.72 = $ Max. Therefore, removing both these spectral components (Fig. 2 inset, green spectrum #2) results in higher peak pulse intensity, $I_{del}^{max}(1;3) = 2.96 I_0$, and slightly shorter pulses (Fig. 2, green line #2). When modulation index $P>3$, three frequency components of the orders 0, 1, and 3 have significant amplitudes and are antiphase accordnig to the respective Bessel functions (Fig. 1). Their removal (Fig. 2 inset, blue spectrum #3) leads to formation of more intense and shorter pulses. Value $P=3.24$ maximizes the sum $-J_{-3}(3.24) - J_{-1}(3.24) - J_0(3.24) = 0.93 = $ Max such that the maximum peak pulse intensity, according to (10), is $I_{del}^{max}(0;1;3) = 3.7 I_0$ (Fig. 2, blue dashed line #3). Higher modulation indices allow forming more intensive and shorter pulses via deletion of larger quantity of spectral components. However, this (i) reduces the average intensity of the pulses because of removal of more and more energy from the field, (ii) increases heights of bursts between the pulses because of non-equidistance of the resulting spectrum, and (iii) can increase the pedestal of the pulse sequence because of non uniform distribution of energy over the spectrum.

Deletion of arbitrary spectral components at arbitrary modulation index results in complicated temporal dependence of the intensity. According to (10), the temporal dependence of intensity is periodic with period of modulation of the field, $2\pi/\Omega$. Within the period, there are oscillations of intensity with various durations and amplitudes. This is the result of interference of the intensity oscillations at harmonics of vibration frequency, $\Omega_m = m\Omega$ ($m=a,b,...$), determined by the removed spectral components (the second group of $N$ items in (8)), and oscillations at combination frequencies of these harmonics (the last group of $C_N^2$ items in (8)). The frequencies of harmonics, $\Omega_m$, are modulated with vibration frequency $\Omega$ and modulation index $P$, $\Omega_m(\tau) = m\Omega + P\Omega\cos(\Omega\tau)$. At sufficiently large modulation index, $P \geq m$, there are time intervals within the period $2\pi/\Omega$ when $\Omega_m \leq 0$. Such an uncommonly large frequency modulation of the intensity oscillations can lead to the temporal dependence of intensity in the form of pulse sequence. Let us consider this regime in the simplest case where some $a$-th single spectral component is deleted at arbitrary modulation index.

In this case intensity (8) takes the form

$$I_{del}(a,\tau) = I_0 \left\{ 1 + J_{-a}^2(P) - 2J_{-a}(P)\cos\left[P\sin(\Omega\tau) + a\Omega\tau\right] \right\}. \tag{11}$$

The instantaneous frequency of intensity oscillations, $\Omega_a(\tau) = a\Omega + P\Omega\cos(\Omega\tau)$, is the $a$-th order harmonic of vibration frequency, $a\Omega$, sinusoidally modulated with modulation index $P$ and vibration period $T = 2\pi/\Omega$. In the case where the modulation index is less than the order "$a$" of the removed spectral component, $P < a$, there are "$a$" oscillations of intensity within the period $T$. The intensity oscillations look like a monochromatic signal with single-tone frequency modulation. They have the same maxima $I_{del}^{max}(a) = I_0 \left\{ 1 + |J_a(P)| \right\}^2$, and minima, $I_{del}^{min}(a) = I_0 \left\{ 1 - |J_a(P)| \right\}^2$. They are the shortest and tightest at the beginning and at the end of period $T$, while they are elongated and more separated in time at the middle of period $T$. In the case where $P=a$, there are still "$a$" oscillations of intensity within the period $T$. However, the instantaneous frequency of oscillations, $\Omega_a(\tau)$, being equal to $2a\Omega$ at the beginning and at the end of period $T$, takes zero magnitude at the middle of the period, $\Omega_a(\pi/\Omega) = 0$. At the middle of period $T$ the phase of intensity oscillations equals $a\pi$ (Fig. 3a, blue line #1), which ensures the lowest value of intensity, $I_{del}(a,\pi/\Omega) = I_0 \left\{ 1 - |J_a(a)| \right\}^2$ and the largest temporal interval between the nearby crests of intensity. As a whole, such oscillations look like bunches of pulses (Fig. 3b, blue line #1). In the case $P > a$, the instantaneous frequency of oscillations, $\Omega_a(\tau)$, is the sign-changing value. Therefore, the phase of intensity oscillations, $\Phi(\tau) = P\sin(\Omega\tau) + a\Omega\tau$, three times takes the magnitude $a\pi$ (Fig. 3a, red line #2). This means that a pair of new bursts (absent in the case $P \leq a$) appear in the middle of period $T$. If the modulation index slightly exceeds the number of the deleted component, $P = a + \xi$ (where $\xi \ll 1$), height of the new bursts is small ($I_{del}^{max} - I_{del}^{min} \approx 4I_0 |J_a| a\xi$). However, these small bursts enlarge the time interval between bunches of the strong oscillations and make them shorter (Fig. 3b, red line #2). It should be noted that for an arbitrary number $a>0$, the Bessel function $J_a(P)$ takes the first maximum at the modulation index just slightly exceeding its order (Fig. 1). This is the reason why deletion of the component with maximum amplitude from the field spectrum (3) leads to formation, within the vibration period, of a single (if $a=1$) or several (if $a>1$) pulses grouped into bunches with $N=a$ pulses in a bunch [35, 36]. Bunches are separated by a pair of relatively small bursts (Fig. 2 black line #1 and Fig. 3b red line #2). Increase of the modulation index leads to increase of amplitude of these bursts up to the amplitude of the bunched strong oscillations,

$I_{del}^{\max}(a) = I_0 \{1 + |J_a(P)|\}^2$ when the phase $\Phi(\tau)$ achieves the magnitude $(a+1)\pi$. Further increase of the modulation index leads to appearance of the next pair of bursts and appearance of the intermediate bunch of strong oscillations of intensity due to the negative magnitudes of the instantaneous frequency, $\Omega_a(\tau)$, (Fig. 3a) and so on, such that in the case $P \gg a$, there is a large number of the intensity oscillations within the period, $T$ (Fig. 3, green lines #3). Their depth, $I_{del}^{\max} - I_{del}^{\min} = 4I_0|J_a|$, is quite small because of small value of the Bessel function, $|J_a(P)| \approx \sqrt{2/\pi P}$ and their quantity, $N$, inside the period, $T$, is determined by the modulation index, $N = \lfloor 2P/\pi \rfloor$ (where the brackets mean the floor function). If $P = \pi N/2 + \xi$ (where $0 < \xi \ll 1$), then amplitude of the $N+1$-th oscillation is small, $I_{del}^{\max} - I_{del}^{\min} \approx I_0|J_a|\xi^2$. As a result, the oscillations of intensity look like groups of pulses (Fig. 3, green lines #3).

Removal of the spectral components with indices $a,b,...$ from the frequency-modulated γ–radiation (3), emitted by the vibrating source, may be implemented by sequentially arranged along z-axis resonant recoilless absorbers "$A,B,...$", moving along z-axis with individually tuned z-projections of the constant velocities, $V_z(A)$, $V_z(B)$,... The resonant frequency of the moving absorber "$A$" is shifted because of the Doppler effect, $\omega_A = \omega_{A0}(1 - V_z(A)/c)$ (where $\omega_{A0}$ is the resonant frequency of the stationary absorber, $c$ is the speed of light). The proper magnitude of $V_z(A)$ provides coincidence of the absorber resonant frequency with frequency of the $a$-th spectral component of the incident radiation (3). In this case, the field of the $a$-th spectral component at the exit of the absorber has the form

$$E_a(\tau) = \tilde{E}_a e^{-i(\omega_a \tau + \Phi_a)} e^{-\Lambda_a/2}, \qquad (12)$$

where

$$\Phi_a = \frac{T_A}{2} \frac{(\omega_a - \omega_A)/\gamma_A}{1 + (\omega_a - \omega_A)^2/\gamma_A^2} \text{ and } \Lambda_a = T_A \frac{1}{1 + (\omega_a - \omega_A)^2/\gamma_A^2} \qquad (13)$$

is the phase incursion due to the resonant dispersion and the exponent of the intensity attenuation due to the resonant absorption at the frequency $\omega_a$ of the $a$-th spectral component, respectively,

$T_A = f_A \dfrac{4\pi \omega_A N_A |d_A|^2 L_A}{\gamma_A \hbar c}$ is the optical thickness of the absorber "$A$" at its transition frequency $\omega_A$ called Mössbauer thickness of the absorber, $f_A$ is the Lamb-Mössbauer factor (the ratio of recoilless to total nuclear resonant absorption), $N_A$ is the density of the resonant nuclei, $d_A$ is the matrix element of the of dipole moment of the corresponding quantum transition, $L_A$ is the thickness of the absorber, $\gamma_A$ is the halfwidth of the spectral line of the resonant transition. Let the frequency of spectral line of the stationary source equals the frequency of spectral line of the stationary absorber, $\omega_{A0} = \omega_0$, then moving the absorber with velocity $V_z(A) = -ac\Omega/\omega_0$ tunes the frequency of its spectral line to the frequency of the $a$-th sideband of the field (3), (5), $\omega_a - \omega_A = 0$. In this case $\Phi_a = 0$, $\Lambda_a = T_A$. The selective filtering the $a$-th sideband occurs if the influence of the absorber on adjacent sidebands is negligible. This condition can be fulfilled if (i) the modulation frequency essentially exceeds halfwidth of the spectral line of the absorber ($\Omega \gg \gamma_A$) and (ii) the following relations are valid,

$$\Lambda_{a\pm 1} \cong T_A (\gamma_A/\Omega)^2 \ll 1, \ |\Phi_{a\pm 1}| \cong T_A \gamma_A/\Omega \ll 1. \qquad (14)$$

Let us consider the conditions similar to the conditions experimentally realized in [35]. Let the radioactive source $^{57}$Co emitting 14.4 keV photons vibrates with frequency $\Omega/2\pi = 10$ MHz providing modulation index $P = 1.84$. Let the width of the resonant spectral line of the absorber is $2\gamma_A/2\pi = 1$ MHz and its Mössbauer thickness is $T_A = 5$ which is realized in a stainless steel foil of thickness $L_A = 25$ μm with natural abundance of nuclide $^{57}$Fe (~2%) [35]. Let the frequency of the resonant spectral line of the absorber is tuned to the frequency of the first sideband of the emitted field, $\omega_A = \omega_0 + \Omega$, via its movement with velocity $V_z(A) = -0.96$mm/s. Then, according to (13), the intensity of the first sideband at the exit of the absorber is $I_1^{out} = I_1^{inp} e^{-\Lambda_1} = I_1^{inp} e^{-T_A} = 0,007 I_1^{inp}$ (where $I_1^{inp}$ is the intensity of the first sideband at the entrance into the absorber). For the nearest zeroth and the second spectral components optical thickness of the absorber is $\Lambda_0 = \Lambda_2 \approx T_A \gamma_A^2 / \Omega^2 = 0.0125$, which provides the output intensities $I_0^{out} / I_0^{inp} = I_2^{out} / I_2^{inp} = 0.99$ and the phase shifts $|\Phi_0| = |\Phi_2| \approx T_A \gamma_A / \Omega = 0.08\pi$. Thus, deletion of the first spectral component of the frequency-modulated γ-radiation by recoilless resonant absorber under the above conditions is selective with high accuracy. Such a deletion leads to formation of a sequence of pulses plotted in Fig. 2 black line #1 which are similar to pulses in [35]. Use of one more stainless steel foil "B" with the same as "A" parameters, moving with velocity $V_z(B) = -2.88$mm/s would provide the selective deletion of the third spectral sideband. At the modulation index $P=2.3$, this would lead to increase of the peak pulse intensity as discussed above (Fig. 2 green line #2). Similarly, three stainless steel foils with velocities $V_z(0) = 0$, $V_z(A) = -0.96$mm/c, and $V_z(B) = -2.88$mm/c at the modulation index $P=3.24$ selectively delete three spectral components further increasing the peak pulse intensity and shortening the pulse duration similar to Fig. 2 blue dashed line #3.

The propagation of γ-radiation through the absorbers is accompanied by the non-resonant attenuation of radiation as a whole owing to photoelectric effect. Intensity of the field behind the absorbers with taking into account the photoelectric absorption can be written as

$$I_{tr}^{pha} = I_{tr} e^{-\mu_A L_A - \mu_B L_B - \cdots}, \qquad (15)$$

where $I_{tr}$ is the intensity of the transformed field, (6), while $\mu_Z$ and $L_Z$ ($Z=A,B,…$) are, respectively, the linear photoelectric absorption coefficient and thickness of the corresponding absorber. The reduction of intensity in an absorber due to photoelectric absorption for a given magnitude of the Mössbauer thickness, $T_A$, is ultimately determined by the percentage of the resonant nuclei: the higher percentage of the resonant nuclei, the smaller thickness, $L_A$, of the absorber and, hence, the smaller photoelectric attenuation. The minimum photoelectric attenuation of intensity occurs in the case where all atoms of an absorber contain the resonant nuclei. For example, in the ideal case of absorber consisting of atoms with $^{57}$Fe nuclei only, the linear photoelectric absorption coefficient is $\mu = 4.95 \times 10^4 m^{-1}$ [50]. This means that the Mössbauer thickness $T_A = 5$ corresponds to $L_A = 0.35 \mu m$ leading to very weak photoelectric attenuation of intensity in the absorber, $I_{del}^{pha} = I_{del} e^{-0.018} = 0.98 I_{del}$ (where $I_{del}^{pha}$ is the intensity of the transformed radiation obtained via deletion of a single spectral component with taking into account the photoelectric absorption). However, in the case of stainless steel foil of the chemical compound Fe:Cr:Ni at 70:19:11 wt% (Type 304), and natural abundance of nuclide $^{57}$Fe (~2%), the Mössbauer thickness $T_A = 5$ corresponds to the foil thickness $L_A = 25$ μm. The linear photoelectric absorption coefficient in this case is $\mu = 5.06 \div 5.17 \times 10^4 m^{-1}$ [50]. As a result, one has strong photoelectric attenuation of intensity in the absorber up to the value $I_{del}^{pha} = I_{del} e^{-1.3} = 0.27 I_{del}$.

There is an option to selectively delete two spectral components of frequency-modulated γ-radiation emitted by a vibrating source by means of using a single resonant absorber with split spectral line. This option can be implemented at the specific magnitude of vibration frequency equal to the frequency splitting of the absorber spectral line. For example, the absorber $FeC_2O_4 \cdot 2H_2O$ has the spectral line, split at ≈17.9MHz [21]. It could be used for simultaneous selective deletion of the first and third sidebands subject to the vibration frequency of 8.95MHz, or for the deletion of the adjacent zero and first sidebands subject to the vibration frequency of 17.9MHz. Similarly, one can use the absorber $FeTiO_3$ with split at 6.98MHz spectral line or the absorber $Fe_2(SO_4)_3 \cdot xH_2O$ with split at 5.6MHz spectral line [21].

In the case where the pulse sequence is formed in a vibrating absorber, (4), from radiation emitted by a stationary source, all the above discussions and formulae (7)-(14) are valid if they are referred to the reference frame, comoving with the absorber. After coming back to the laboratory reference frame, the field passed across the vibrating absorber has the form differing from (7) by the factor $e^{-iP\sin(\Omega\tau+\vartheta_0)}$,

$$E_{del}^{lab}(a,b,...,\vartheta_0,\tau) = e^{-iP\sin(\Omega\tau+\vartheta_0)} \sum_{n=-\infty}^{\infty} \tilde{E}_n e^{-i\omega_n\tau} \underbrace{-\tilde{E}_a e^{-i\omega_a\tau} - \tilde{E}_b e^{-i\omega_b\tau} - ...}_{N}, \quad (16)$$

whereas the intensity keeps the form (8). The transformation of recoilless γ–radiation into the pulse sequence in a single vibrating absorber has been realized in [35]. In that case, tuning the spectral line of absorber to the definite sideband of radiation was realized via moving the source with a constant velocity along z-direction. In order to use several vibrating absorbers for the selective deletion of several spectral components one should provide synchronization of vibrations of all the absorbers with the same frequency and their movement with individual constant velocities along z-direction.

### IV. PHASE INVERSION OF SPECTRAL COMPONENTS

As noted in section II, the spectrum (5) of the frequency-modulated radiation, (3), consists of in-phase and antiphase components having the initial phase of $0 - n\vartheta_0$ and $\pi - n\vartheta_0$, respectively. Therefore, the selective inversion (i.e., shift by π) of the initial phases of either antiphase or in-phase components using the resonant dispersion of the properly tuned absorbers would result in the phase-aligned spectrum corresponding to a periodic sequence of bandwidth-limited pulses. Their peak intensity would be essentially higher while pedestal and bursts between the pulses essentially smaller compared to the case of deletion of the mismatched spectral components since the energy is not removed from the field and the spectrum remains equidistant.

Inversion of initial phases of N arbitrary spectral components of the spectrum (5) with numbers $m = \underbrace{a,b,c,...}_{N}$ corresponds to $\tilde{\eta}_m = -\eta_m$, where $\eta_m = e^{-\Lambda_m/2}$ in (6). The resulting field, $E_{inv}$, and its intensity, $I_{inv} = c|E_{inv}|^2/8\pi$, are, respectively,

$$E_{inv}(a,b,...,\vartheta_0,\tau) = \sum_{n=-\infty}^{\infty} \tilde{E}_n e^{-i\omega_n\tau} \underbrace{-(1+\eta_a)\tilde{E}_a e^{-i\omega_a\tau} -(1+\eta_b)\tilde{E}_b e^{-i\omega_b\tau} - ...}_{N}, \quad (17)$$

and

$$I_{inv}(a,b,...,\vartheta_0,\tau) = I_0 \left\{ 1 + \underbrace{G_{-a}^2 + G_{-b}^2 + ...}_{N} \right.$$

$$\underbrace{-2G_{-a}\cos(P\sin\alpha + a\alpha) - 2G_{-b}\cos(P\sin\alpha + b\alpha) - ...}_{N}$$

$$\underbrace{+2G_{-a}G_{-b}\cos((a-b)\alpha) + 2G_{-a}G_{-c}\cos((a-c)\alpha) + 2G_{-c}G_{-b}\cos((c-b)\alpha) + ...}_{C_N^2} \quad (18)$$

where $G_m = (1+\eta_m)J_m$. According to (18), inversion of initial phases of N arbitrary components with numbers $\underbrace{a,b,c,...}_{N}$ of the spectrum (5) of the frequency-modulated field (3) results in the same temporal dependence of the intensity, as inversion of initial phases of N components with numbers $\underbrace{-a,-b,-c,...}_{N}$ (symmetrically located with respect to the carrier frequency $\omega_0$) subject to the initial phase $\vartheta_0$ is shifted by π. In other words, the following relation is valid,

$$I_{inv}(a,b,...,\vartheta_0,\tau) = I_{inv}(-a,-b,...,\vartheta_0 \pm \pi,\tau). \tag{19}$$

Therefore, similar to the previous paragraph, we assume below $\vartheta_0 = 0$ and consider inversion of initial phases of antiphase components only.

Comparison of formulae (7)-(9) with formulae (17)-(19) shows that the selective phase inversion is formally similar to the case of deletion of the spectral components. Similar to (10), the highest peak pulse intensity is achieved at $\tau=0$ and equals

$$I_{Inv}(a,b,...,\tau = 0) = I_0 \left( 1\underbrace{-G_{-a}-G_{-b}...}_{N} \right)^2. \tag{20}$$

The peak pulse intensity takes its maximum value, $I_{inv}(a,b,...,\tau = 0) = \max\{I_{inv}(a,b,...)\}$, in the case where all the antiphase spectral components with noticeable amplitudes are inverted and the vibration amplitude provides the optimal value of modulation index, $P$, such that sum $-G_{-a}-G_{-b}...$ takes the largest magnitude. As follows from (10) and (20), inversion of initial phases of antiphase components allows one to increase the maximum peak pulse intensity less than four times, $\max\{I_{inv}(a,b,...)\}/\max\{I_{del}(a,b,...)\} < 4$, compared to the case of deletion of these components at the same parameter magnitudes and under the ideal conditions where the only change in the spectrum (5) is either phase inversion or deletion of the components. As discussed below, this ratio is reduced depending on the possible implementations and unaccounted yet factors.

The resonant dispersion is inevitably accompanied by the resonant absorption. Therefore the optimal conditions for the selective resonant inversion of initial phase of the *a*-th spectral component should imply a proper detuning of the absorber resonant frequency from the frequency of the *a*-th component, $\delta_a = \omega_a - \omega_A$, in order to minimize attenuation of this component and to minimally alter the adjacent components. Let $\delta_a > 0$, then as follows from (13), larger detuning leads to weaker resonant absorption of the *a*-th spectral component but stronger influence of the absorber resonant transition on the nearest *a*−1-th component. For example, let the vibration frequency is $\Omega = 20\gamma_A$ (corresponding to $\Omega/2\pi = 10$ MHz in the experiment with $^{57}$Fe, [35]) and the detuning is $\delta_a = 3\gamma_A$. Then, according to (13), inversion of initial phase of the *a*-th spectral component occurs at Mössbauer thickness $T_A = 20.9$ leading to both strong resonant attenuation of this component (its intensity at the exit of the absorber is $I_a^{out} = 0{,}12 I_a^{inp}$, Fig. 4a) and noticeable change of the *a*−1-th component (the phase incursion is $\Phi_{n-1} = -0.2\pi$ and the output intensity is $I_{a-1}^{out} = 0{,}93 I_{a-1}^{inp}$). This means that inversion of initial phase of the first spectral component or both the first and third components in one or two resonant absorbers, respectively, under the conditions discussed in Sec.III, has no significant advantage over the selective resonant deletion of these components. Moreover, the large Mössbauer thickness of the absorber ($T_A = 20.9$) means noticeable reduction of intensity as a whole due to photoelectric absorption, especially if two absorbers are used. However, in the case $\Omega \gg \delta_a \gg \gamma_A$, the resonant dispersion can provide selective inversion of initial phase of the *a*-th component since, according to (13), one has $\Lambda_a \approx 2\pi\gamma_A / \delta_a \ll 1$, $\Lambda_{a-1} \approx \Lambda_a (\delta_a/\Omega)^2 \ll 1$, and $|\Phi_{a-1}| \approx |-\pi\delta_a/\Omega| \ll 1$, subject to sufficiently large Mössbauer thickness of absorber, $T_A \approx 2\pi\delta_a / \gamma_A \gg 1$. For instance, if $\Omega = 100\gamma_A$ (corresponding to $\Omega/2\pi = 50$ MHz in the case of $^{57}$Fe, [35]) and $\delta_a = 15\gamma_A$, one has $\Lambda_a = 0.4$, $\Lambda_{a-1} = 0{,}009$, $|\Phi_{a-1}| = 0.15\pi$, and $T_A = 94.3$, which corresponds to rather weak resonant attenuation of the phase inverted *a*-th component ($I_a^{out} = 0.67 I_a^{inp}$) and small resonant alteration of the adjacent components. However, even in the case of the ideal absorber consisting of only atoms with $^{57}$Fe nuclei, this requires the absorber thickness $L_A=6.6\mu m$ resulting, according to (15), in about 30% reduction of radiation intensity due to photoelectric absorption.

Use of the resonant dispersion can be useful when the modulation index, $P$, is such that two adjacent spectral components are antiphase. Such a situation occurs at $P>2.4$. In this case, the resonant spectral line of a single absorber can be tuned in the middle between these components (Fig. 4b). If the Mössbauer thickness of absorber is $T_A=\frac{4\pi\gamma_A}{\Omega}\left(1+(\Omega/2\gamma_A)^2\right)$, then, according to (13), at the exit of the absorber initial phase of the $a$-th spectral component is shifted by $\pi$ while initial phase of the $a-1$-th spectral component is shifted by $-\pi$. Thus, a resonant absorber with unsplit spectral line simultaneously inverts initial phases of two spectral components. The resonant attenuation of both components is the same, $\Lambda_{a-1}=\Lambda_a=4\pi\gamma_A/\Omega$. The alteration of other spectral components in the case $\Omega\gg\gamma_A$, is defined by the relations: $\Lambda_{a-1-m}=\Lambda_{a+m}\approx\Lambda_a/(2m+1)^2$, $|\Phi_{a-1-m}|=|\Phi_{a+m}|\approx\pi/(2m+1)$, $m=1,2,\ldots$ Thus, simultaneous inversion of initial phases of two spectral components in a single absorber instead of their deletion by two absorbers enables one (i) to transform the same spectrum with fewer resonant absorbers and (ii) to save two spectral components in the spectrum. The latter reduces both the pedestal of the pulse sequence and heights of bursts between the pulses, as well as increases the peak pulse intensity. However, (i) the components nearest to the phase-inverted ones acquire rather large phase shifts of $\pm\pi/3$ for $m=\pm1$, and $\pm\pi/5$ for $m=\pm2$ at the exit of the absorber, leading to lengthening of the pulses, and (ii) the required Mössbauer thickness of the absorber is quite large (the minimum value of $T_A$ is $T_A^{\min}=8\pi\approx25$, realized at $\Omega=2\gamma_A$, while in the case $\Omega\gg\gamma_A$, one has $T_A\approx\pi\Omega/\gamma_A\gg1$), which reduces the output intensity because of the photoelectric absorption.

For example, in the case where vibration frequency is $\Omega=20\gamma_A$ and modulation index is $P=3.24$ (Fig. 2, blue dashed line and spectrum in inset #3), one can use a single absorber with Mössbauer thickness $T_A=63.5$ to invert initial phases of two spectral components #0 and #1 instead of using two absorbers with $T_A=5$ for their deletion (Fig. 5a). The exponent of the accompanied resonant absorption is $\Lambda_0=\Lambda_1=0.63$ corresponding to the output intensity of these components $I_0^{out}/I_0^{inp}=I_1^{out}/I_1^{inp}=0.53$. The nearest minus first and second components acquire the phase shift of $\pm\pi/3$ and are resonantly attenuated such that $I_{-1}^{out}/I_{-1}^{inp}=I_2^{out}/I_2^{inp}=0.93$. Amplitude and phase alteration of the minus first and second components in addition to initial phase inversion of #0 and #1 and deletion of #3 components, leads to modification of relations (10) and (20) such that the highest peak pulse intensity is achieved at optimal modulation index $P=3.4$. The required Mössbauer thickness $T_A=63.5$ in the case of ideal $^{57}$Fe absorber corresponds to the absorber thickness $L_A=4.4\mu m$ leading, according to (15), to 20% reduction of intensity as a whole, $I_{inv}^{pha}=I_{inv}e^{-0.2}=0.8I_{inv}$ (where $I_{inv}^{pha}$ and $I_{inv}$ is the intensity of radiation after inversion of initial phases of two spectral components with and without photoelectric absorption, respectively) versus 4% reduction of intensity in two absorbers deleting these components. As a result, in the case of optimal modulation index, $P=3.4$, use of the ideal absorber "$A$" with thickness $L_A=4.4\mu m$ inverting initial phases of zero and the first spectral components, and sequentially placed absorber "$B$" with thickness $L_B=0.35\mu m$ deleting the third component (Fig. 5a), leads to pulses with peak intensity less than peak intensity of pulses produced via deletion of all three components at optimal modulation index $P=3.24$ (Fig. 5b). However, the pedestal of the pulse sequence produced via phase inversion of the anti-phase spectral components is lower than the pedestal of the pulse sequence produced via deletion of these components owing to more uniform distribution of energy over the spectrum in the former case. At higher magnitudes of the modulation index, $P$, optimal for pulse formation via deletion or inversion of the antiphase components, these differences also take place. If the ideal absorbers are replaced with compounds (like stainless steel foils), the difference in peak pulse intensity and pedestal becomes larger since the foil thicknesses and hence photoelectric absorption are increased with decreasing percentage of $^{57}$Fe in compound. Thus, in the case of $^{57}$Co source of γ-radiation and $^{57}$Fe absorbing nuclei, use of $N$ ($N≥3$) resonant absorbers for deleting the set of $N$ antiphase spectral components from the spectrum of frequency-modulated radiation results in a periodic sequence of pulses with the same duration but larger peak intensity, pedestal, and bursts between the pulses as compared with use of $N−1$ resonant absorbers, one of which inverts phases of two spectral components of this set.

However, use of $N$ ($N≥3$) resonant absorbers, one of which inverts phases of two adjacent spectral components, enables one to transform $N+1$ antiphase spectral components of the frequency-modulated radiation at higher modulation index as compared with use of $N$ resonant absorbers only deleting the spectral components. This can result in shorter pulses with higher peak intensity. For example, in the case where vibration frequency is $\Omega=20\gamma_A$ and modulation index is $P=4.7$, four spectral components are antiphase (Fig. 6a). In this case, three resonant absorbers are sufficient for their alteration: the absorber "A" with Mössbauer thickness $T_A=63.5$ inverts the components #−1 and #0, while the absorbers "B" and "C" with Mössbauer thickness $T_B=T_C=5$ delete the spectral components #3 and #5. This results in shorter pulses with higher peak intensity, using both ideal and stainless steel absorbers, as compared to the case of the same quantity of resonant absorbers with Mössbauer thickness $T_A=5$ deleting three spectral components at $P=3.24$ (Fig. 6b). The peak intensity of the produced pulses is more than for times higher than intensity of γ-radiation from the source, while the pulse duration is less than one tenth of the repetition period.

At larger magnitudes of the modulation index several pairs of adjacent antiphase spectral components can occur. As seen from Fig.1, pairs of spectral components of the following orders are antiphase: {-1,-2} and {2,3} for $P≈6$ since $J_1,J_2$ and $J_{-2},J_{-3}$ are negative, {-2,-3} and {1,2} for $P≈7.3$ since $J_2,J_3$ and $J_{-1},J_{-2}$ are negative, {-2,-3}, {1,2}, and {4,5} for $P≈8$ since $J_2,J_3$, $J_{-1},J_{-2}$, and $J_{-4},J_{-5}$ are negative. Each pair can be phase inverted by one respective absorber. Thus, use of the resonant dispersion allows one to obtain the widest spectrum of nearly phase-aligned equidistant components corresponding to the shortest pulses using minimum quantity of absorbers. For example, in the case of $P≈8$, the pulse duration is evaluated as 0.05 of the repetition period that corresponds to 5 ns for the vibration period of 10 MHz. However, the photoelectric absorption essentially reduces the intensity of γ-radiation as a whole in each dispersive absorber compared to the reduction of intensity in resonant absorbers selectively deleting the same components. Combining phase inversion of some components and deletion of other components allows one to produce periodic sequence of ultrashort γ-ray pulses with optimal peak intensity and properly reduced intermediate bursts and pedestal.

## V CONCLUSION

In this paper, we studied the possible ways to develop the technique [35] of shaping recoilless γ-radiation of a radioactive source into a periodic sequence of ultrashort pulses via frequency modulation of radiation and transformation of its spectrum in a resonant absorbing medium. The possibilities to produce shorter γ-ray pulses with higher peak intensity were investigated. Our study was based on an analytical model assuming γ-radiation as a monochromatic wave with sinusoidally modulated carrier frequency. The frequency modulation of γ-radiation is provided by mechanical sinusoidal vibration of either a source or absorbing medium along the direction of radiation propagation. The resonant absorbing medium is used to transform the specific spectrum (consisting of in-phase and antiphase components) of the frequency-modulated radiation with time-independent intensity into a spectrum of phase-aligned components corresponding to a periodic sequence of pulses. The shortening of pulses and increase of their peak intensity is achieved via increase of the modulation index, which leads to appearance of more and more in-phase and antiphase components. We proposed to correct the phase mismatching between the spectral components and build the phase-aligned spectrum by means of using a set of several sequentially placed resonant absorbers. The two basic techniques were discussed.

The first technique is based on the selective deletion of all non-vanishing mismatched spectral components via their resonant absorption in the spectrally tuned absorbers such that the remaining components form the phase-aligned spectrum. The resulting pulses are nearly bandwidth limited. However, the pulse sequence can have a noticeable pedestal and rather intensive bursts between the pulses since the spectrum becomes non-equidistant after deletion of the spectral components. This technique can be realized with the modest Mössbauer thickness of the absorbers providing weak nonresonant attenuation of radiation as a whole due to photoelectric absorption. As a result the peak pulse intensity can essentially exceed the intensity of radiation emitted by the source. The required quantity of the absorbers equals the quantity of the deleted spectral components. This factor can impede to shorten the pulses because of the technical problems to manage a large number of the absorbers. The deletion of more and more spectral components increases non-equidistance of the spectrum and reduces the average intensity of the pulses.

We also studied the selective deletion of arbitrary single spectral component from the spectrum of the frequency modulated γ-radiation and showed that there exist the optimal conditions when deleting a single spectral component of arbitrary order, $m$, results in a periodic sequence of pulse bunches containing $m$ pules in a bunch.

Another technique is based on selective inversion of initial phases of the mismatched spectral components using the resonant dispersion of absorbers. This technique enables building the equidistant spectrum of nearly aligned components. If the resonant frequency of an absorber is tuned in the middle between two adjacent spectral components, a single absorber can simultaneously phase-invert them both, reducing the number of absorbers required to build wide phase aligned spectrum. Hence, potentially, it enables one to produce shorter pulses with fewer absorbers compared to the technique based on deletion of the components. The pulse sequence has lower pedestal and smaller intermediate bursts. However, this technique needs substantially larger Mössbauer thicknesses of the absorbers resulting in essential attenuation of radiation as a whole due to photoelectric absorption. As a result, shortening of pulses with larger number of the absorbers is accompanied by essential reduction of their peak intensity.

We considered the case of $^{57}$Co source of γ-radiation vibrating with frequency 10 MHz and various experimentally achievable amplitudes as well as various numbers of absorbers from one to three with different percentage of $^{57}$Fe resonant nuclei. It was shown that taking into account the advantages of the discussed techniques and combining the dispersive and absorptive properties of the resonant absorbers, one can control peak pulse intensity, pulse duration, and pedestal of the pulse sequence. Use of three resonant absorbers in experimentally realizable conditions allows one to produce pulses with duration about 0.08 of the vibration period and peak intensity more than 4 times exceeding intensity of the source radiation.

### IV. ACKNOWLEDGEMENTS

We acknowledge the support from the US NSF (grant no. PHY-1307346), the RFBR (grants nos 13-02-00831, № 14-02-31044, and 14-02-00762), and The Ministry of Education and Science of the Russian Federation (contract No. 11.G34.31.0011). V.A. acknowledges a personal grant for young scientists from "Dynasty" Foundation. The work is performed according to the Russian Government Program of Competitive Growth of Kazan Federal University.

[*]*Corresponding author: radion@appl.sci-nnov.ru*

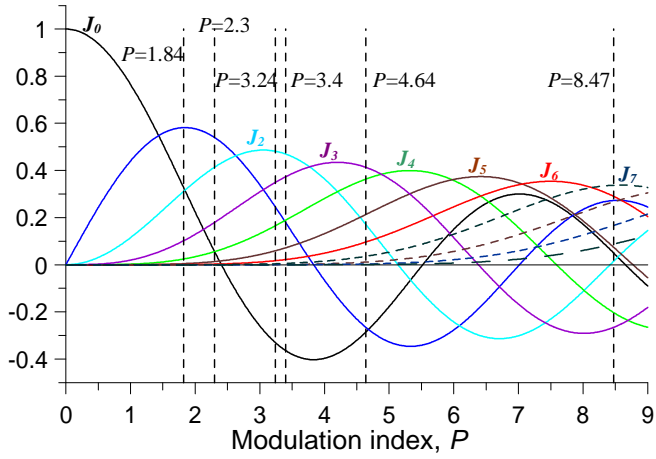

Fig. 1. (Color online) Bessel functions from $J_0(P)$ up to $J_{10}(P)$ versus modulation index, $P$. The dashed not labeled lines starting up from zero more and more gradually, correspond to $J_8(P)$, $J_9(P)$, and $J_{10}(P)$, respectively. Amplitudes and initial phases of the respective spectral components in (5) are determined by the Bessel function order, taking into account the relation $J_{-n} = (-1)^n J_n$

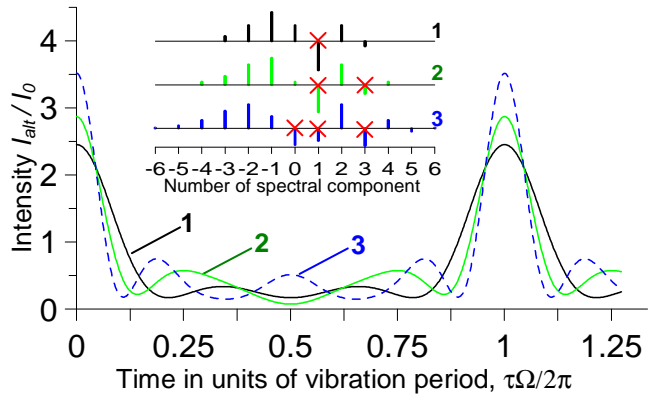

Fig.2. (Color online) Normalized intensity of the transformed γ-radiation, $I_{tr}/I_0$, produced from the frequency modulated radiation, (3), via deletion of the first component from the spectrum (5) at modulation index $P=1.84$ (black line #1 and the corresponding spectrum #1 in inset); the first and third spectral components at modulation index $P=2.3$ (green line #2 and the corresponding spectrum #2 in inset); zero, first and third spectral components at modulation index $P=3.24$ (blue dashed line #3 and the corresponding spectrum #3 in inset) in one, two, and three ideal resonant absorbers, respectively, with $T_A=5$. Intensity of the transformed radiation, $I_{tr}/I_0 = I_{del}^{pha}/I_0$, takes into account the photoelectric absorption in ideal $^{57}$Fe absorbers. In inset, relative amplitudes and initial phases of spectral components of the field (3), (5) are represented by the vertical bars (in arbitrary units). The in-phase components are above the x-axis while the antiphase components are below the x-axis. The deleted spectral components are marked by red criss-crosses.

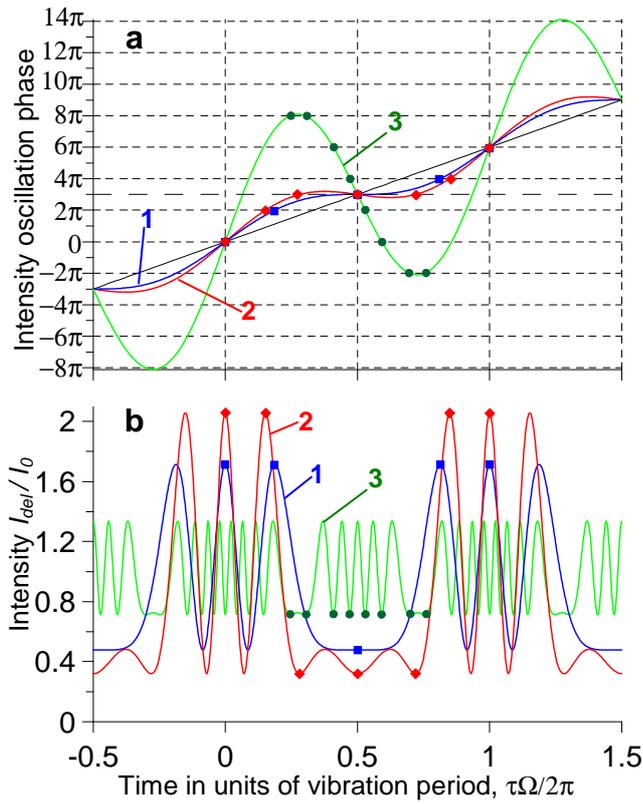

Fig.3. (Color online) **a**: Phase, $\Phi(\tau) = P\sin(\Omega\tau) + a\Omega\tau$, of intensity oscillation, (7), versus time normalized to the vibration period, for the case of the third ($a=3$) deleted spectral component from the spectrum (5) subject to $P=a=3$ (blue line #1), $P=4.2$ (red line #2), $P=20.5$ (green line #3); **b**: the corresponding normalized intensities of γ-radiation, $I_{del}^{pha}/I_0$.

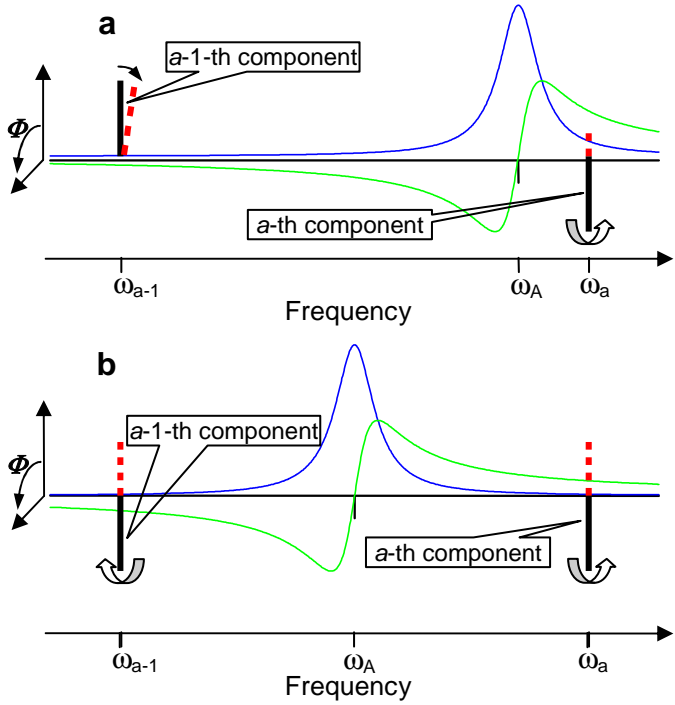

Fig.4. (Color online) Relative amplitudes (vertical bars in arbitrary units) and initial phases, $\Phi$, of the $a$-th and $a$-1-th spectral components of the frequency modulated field (3), (5) before (black bars) and after (red dashed bars) alteration in a resonant absorber. The in-phase components are above the x-axis while the antiphase components are below the x-axis. Green and blue lines represent spectral profiles of the resonant dispersion and absorption, respectively, of an absorber. **a**: Inversion of initial phase of the $a$-th antiphase spectral component via the resonant dispersion of an absorber is accompanied by its attenuation due to the resonant absorption. The nearest $a$-1-th spectral component acquires a phase shift due to the resonant dispersion. The absorber has Mössbauer thickness $T_A=20.9$, its spectral line with halfwidth $\gamma_A$ is detuned from the frequency of the $a$-th spectral component by $\delta_a=3\gamma_A$, the absorber vibrates with frequency $\Omega=20\gamma_A$. **b**: Simultaneous inversion of initial phases of the $a$-th and $a-1$-th antiphase spectral components via the resonant dispersion of absorber with taking into account their attenuation due to the resonant absorption. The Mössbauer thickness of absorber is $T_A=63.5$.

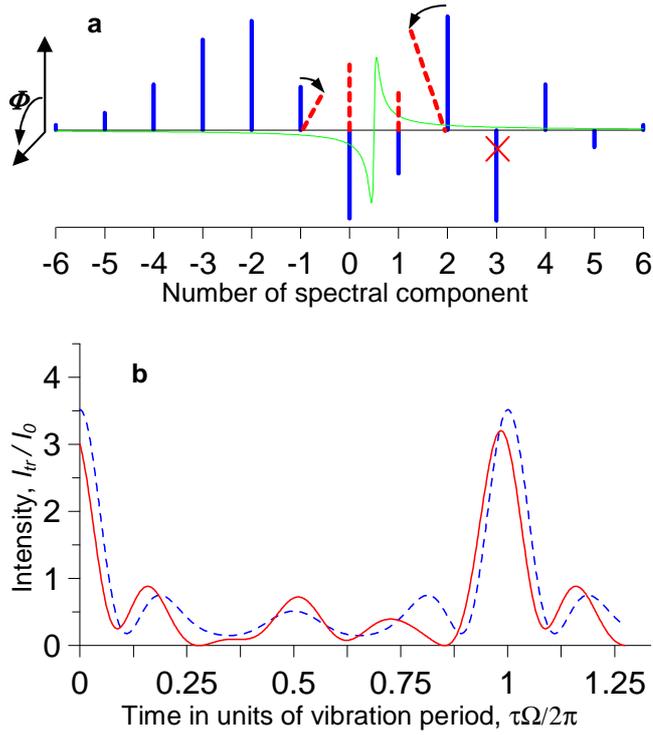

Fig.5. (Color online) **a**: Relative amplitudes (vertical bars in arbitrary units) and initial phases, $\Phi$, of spectral components of the frequency modulated field (3), (5) at optimal modulation index $P=3.4$ and modulation frequency $\Omega=20\gamma_A$ before (blue bars) and after (red dashed bars) alteration in two resonant absorbers (unchanged components are the same blue bars). The in-phase components are above the x-axis while the antiphase components are below the x-axis. One absorber with $T_A=63.5$ inverts initial phases of zero and the first components via its resonant dispersion (green curve) and reduces their amplitudes because of the resonant absorption as well as changes initial phases of the adjacent components by $\pm\pi/3$ and slightly attenuates them. Another absorber with $T_A=5$ deletes the third component (marked by criss-cross). **b:** The normalized intensity of the transformed field, $I_{tr}/I_0 = I_{inv}^{pha}/I_0$, red curve, in comparison with intensity of the field formed in three resonant absorbers deleting all three spectral components at optimal modulation index $P=3.24$, $I_{tr}/I_0 = I_{del}^{pha}/I_0$ (blue dashed line which is the same as blue dashed line #3 in Fig.2). Both curves take into account non-resonant photoelectric absorption of the field as a whole for the case of ideal $^{57}$Fe absorbers.

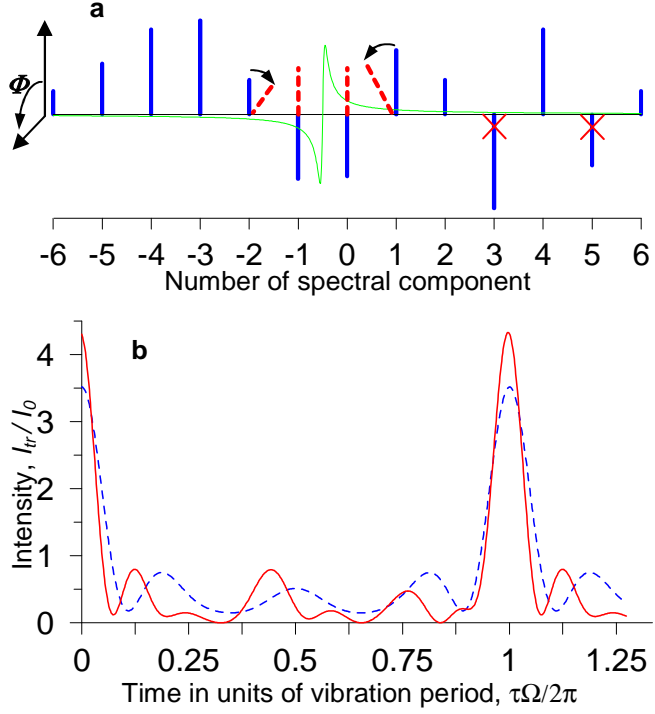

Fig.6. (Color online) **a**: Relative amplitudes (vertical bars in arbitrary units) and initial phases, $\Phi$, of spectral components of the frequency modulated field (3), (5) at optimal modulation index $P=4.7$ and modulation frequency $\Omega=20\gamma_A$ before (blue bars) and after (red dashed bars) alteration in three resonant absorbers (unchanged components are the same blue bars). The in-phase components are above the x-axis while the antiphase components are below the x-axis. One absorber with $T_A=63.5$ inverts initial phases of the minus first and zero components via its resonant dispersion (green curve) and reduces their amplitudes because of the resonant absorption as well as changes initial phases of the adjacent components by $\pm\pi/3$ and slightly attenuates them. Two other absorbers with $T_A=5$ delete the third and fifth components (marked by criss-crosses). **b:** The normalized intensity of the transformed field, $I_{tr}/I_0 = I_{inv}^{pha}/I_0$, red curve, in comparison with intensity of the field formed in three resonant absorbers deleting three spectral components at optimal modulation index $P=3.24$, $I_{tr}/I_0 = I_{del}^{pha}/I_0$ (blue dashed line which is the same as blue dashed line #3 in Fig.2). Both curves take into account non-resonant photoelectric absorption of the field as a whole for the case of ideal $^{57}$Fe absorbers. Peak intensity of red pulses is 1.23 of peak intensity of blue pulses. In the case of stainless steel absorbers fully enriched by $^{57}$Fe, this ratio is 1.13. The red pulse duration is about 0.08 of the vibration period, corresponding to 8 ns in the case where vibration frequency is 10 MHz.